\documentclass[pra, twocolumn, showpacs, floatfix, nobalancelastpage, superscriptaddress]{revtex4-1}
\usepackage{amsmath}
\usepackage{graphicx}
\usepackage{amsfonts}
\usepackage{amssymb}
\usepackage{epstopdf}
\usepackage{bbold}
\usepackage{color}

\DeclareMathAlphabet\mathbfcal{OMS}{cmsy}{b}{n}

\newcommand{\gr}[1]{\boldsymbol{#1}}
\newcommand{\ket}[1]{| #1 \rangle}
\newcommand{\bra}[1]{\langle #1 |}
\newcommand{\braket}[2]{\langle #1 | #2 \rangle }

\newcommand{\tr}{\mathrm{Tr}}
\renewcommand{\t}[1]{\mathrm{#1}}

\begin{document}
\title{Compatibility in multiparameter quantum metrology}

\author{Sammy Ragy}
\email{sammy.ragy@york.ac.uk}
\affiliation{Department of Mathematics, University of York,
Heslington, York YO10 5DD, United Kingdom}

\author{Marcin Jarzyna}
\affiliation{Faculty of Physics, University of Warsaw, ul. Pasteura 5, 02-093 Warszawa, Poland}

\author{Rafa\l \, Demkowicz-Dobrza\'{n}ski}
\affiliation{Faculty of Physics, University of Warsaw, ul. Pasteura 5, 02-093 Warszawa, Poland}

\begin{abstract}
Simultaneous estimation of multiple parameters in quantum metrological models is complicated by factors relating to the (i) existence of a single probe state allowing for optimal sensitivity for all parameters of interest, (ii) existence of a single measurement  optimally extracting information from the probe state on all the parameters, and (iii) statistical independence of the estimated parameters. We consider the situation when these concerns present no obstacle and for \textit{every} estimated parameter the variance obtained in the multiparameter scheme is equal to that of an optimal scheme for that parameter alone, assuming all other parameters are perfectly known. We call such models \textit{compatible}. In establishing a rigorous theoretical framework for investigating compatibility, we clarify some ambiguities and inconsistencies present in the literature and discuss several examples to highlight interesting features of unitary and non-unitary parameter estimation, as well as deriving new bounds for physical problems of interest, such as the simultaneous estimation of phase and local dephasing.
\end{abstract}

\maketitle
\section{Introduction}

The foundations of quantum estimation theory were laid in the sixties and seventies, with the two most significant contributions from Holevo \cite{Holevo} and Helstrom \cite{Helstrom}. Since then the topic has captured the attention of both the physical and mathematical communities.
Most of the activity in the physical community focused on single parameter estimation with particular focus on estimating a unitary parameter, such as phase \cite{MetrologyReview,BraCav, Toth2014, Demkowicz2015}.
In recent years, however, building on existing results on multiple parameter estimation in the mathematical literature \cite{Matsumoto,HayashiAsymptotic,Guta}, there have been a number of theoretical and experimental papers by physicists also addressing the multiple parameter case.
These include estimating multiple-parameter unitary operators \cite{Bagan2001d, Chiribella2004, Kolenderski2008,Genoni,Ballester, two-step, Fujiwara,Multiphase, Berry2015, Baumgratz2016},  estimating  both unitary and decoherence parameters \cite{Durkin,Mesh,Meshloss}, or two decoherence parameters simultaneously \cite{Illuminati}, see \cite{Datta2016} for a short review on the topic.

 Typically, when estimating multiple parameters simultaneously, there is a trade-off in how well different parameters may be estimated. When the estimation protocol is optimized from the point of view of one parameter, the precision of estimating the remaining ones deteriorates. In such cases in order to define a meaningful concept of an optimal multiparameter estimation protocol one e.g. needs to assign weights to different parameters and ask for a protocol minimizing the weighted sum of variances of different parameters.


In this paper, we consider finite-dimensional quantum systems and investigate the conditions when the above mentioned trade-off is not present and
there exists a jointly optimal multiparameter estimation protocol, meaning its performance for each of the parameters matches that of a protocol optimally designed to estimate that parameter assuming all the remaining ones are perfectly known. This essentially results in the maximal advantage over having such separate schemes for each parameter. We choose to call such protocols \emph{compatible}, owing to the fact that a particularly quantum feature of this trade-off occurs in the measurement stage, where it is possible that the optimal measurements for different parameters correspond to incompatible (non-commuting) observables. However, measurement compatibility is but one of several conditions we require for metrological compatibility in general.
\begin{figure}[t]
\includegraphics[width=0.9 \columnwidth]{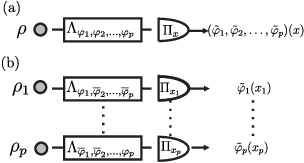}
\caption{(a) Simultaneous estimation of multiple parameters $(\varphi_1,\dots,\varphi_p)$ based on results
of a single measurement performed on the output of a quantum channel $\Lambda_{\varphi_1,\dots,\varphi_p}$ acting on a single input probe $\rho$. (b) $p$ separate schemes where one estimates each parameter individually using dedicated probe states and measurements, treating in every run the remaining parameters as perfectly known. We say that the parameters $(\varphi_1,\dots,\varphi_p)$
of the quantum channel estimation model are compatible if there exists a simultaneous estimation scheme
where each parameter is estimated equally well as in a set of $p$ optimal schemes for the individual parameters, thus leading to a factor $p$ reduction in resources used.}
\label{fig:metrology}
\end{figure}

To get a `like for like' comparison of performance, it is necessary to consider some concept of the resources utilised in a metrological scheme; after all, the variance of an estimation can be made arbitrarily small by simply repeating an experiment to gather more data. For our purposes, we count the number of channel applications. Usually, since we consider single-qubit channels acting in parallel, this will also correspond to the number of qubits in the probe state. This manner of thinking also makes clear the motivation for multiparameter metrology. Should we wish to consume the fewest resources (as might be the case for a channel consisting of a sample fragile to exposure to too many photons), then it may be that we wish to extract the information about all relevant parameters of interest in the same experiment.

The paper is organized as follows.  In Section~\ref{sec:problem} we formulate the framework of multiparameter quantum metrology and
discuss the requirements for \emph{compatible} multiparameter estimation. We also discuss variants of the protocols depending on the use of entanglement at the input as well as at the measurement stages.
In Section~\ref{sec:crbounds} we review the multiparameter classical Cram{\'e}r-Rao (CR) bound, as well as two of its quantum generalizations: the quantum Fisher information (QFI) CR Bound and the Holevo CR bound.
In Section~\ref{sec:saturability} we provide a simple proof for a necessary and sufficient condition for the equivalence of the QFI CR bound with the Holevo CR bound and hence asymptotic saturability of the QFI multiparameter CR bound. In Section~\ref{sec:unitary} we consider a general scheme of multiparameter unitary estimation and provide an explicit structure of generating Hamiltonians
that is necessary and sufficient to satisfy the compatibility requirements. In particular, we prove that when considering simultaneous estimation of angles of rotations of a spin $j$ particle around different axes, the only non-trivial case satisfying the compatibility conditions is the $j=1$ case with the axes of rotation being orthogonal. In Section~\ref{sec:nonunitary}, we turn our attention to the compatible estimation of unitary and decoherence parameters, discussing some sufficient conditions for when this is possible. As an illustration, we analyze in more detail phase estimation in the presence of loss and local dephasing. While symmetric lossy interferometry
is an example of a compatible estimation problem, the local dephasing case manifests incompatibility due to the lack of a single optimal probe
even though all other conditions for simultaneous measurability as well as  statistical independence are satisfied.
 Finally, in Section~\ref{sec:conclusions}, we conclude the paper.

\section{Formulation of the problem}
\label{sec:problem}

Let $\Lambda_{\gr{\varphi}}$ be a quantum channel depending on a set of parameters $\gr{\varphi}=(\varphi_1,\dots,\varphi_p)$
that we want to estimate by sending an input quantum probe $\rho$ and measuring the output $\rho_{\gr{\varphi}} = \Lambda_{\gr{\varphi}}(\rho)$
with a general measurement $\{\Pi_{\gr{x}}\}$. Measurement results are distributed according to a probability distribution $p(\gr{x}|\gr{\varphi}) =
\tr(\rho_{\gr{\varphi}}  \Pi_{\gr{x}})$ and based on their values parameters are estimated using an estimator function $\tilde{\gr{\varphi}}(\gr{x})=(\tilde{\varphi}_1,\dots,\tilde{\varphi}_p)(\gr{x})$, see Fig.~\ref{fig:metrology} (a).
Clearly, estimating multiple channel parameters simultaneously in a single estimation scheme is in general more challenging
than estimating each of the parameters separately using dedicated schemes as in Fig.~\ref{fig:metrology} (b). When estimating each parameter separately one is entitled to choose a probe state and a measurement
which are optimal for enhancing the sensitivity of the scheme with respect to this particular parameter.

Still, a simultaneous metrology scheme may sometimes match the performance of the separate schemes (while using only the resources of one of them) provided the three following conditions are satisfied:
(i) there is a single probe state $\rho$ with which one can replace all input states $\rho_i$ in the separate schemes
preserving the maximal sensitivity of the output probe with respect to all the parameters,
 (ii) there is a single measurement $\{\Pi_{\gr{x}}\}$ (where $\gr{x}$ will generally be a vector of data) that can replace all
 measurements $\{\Pi_{x_i}\}$ in the separate schemes and yield optimal precision for each parameter, and finally (iii)
under requirement of preserving optimal precision for estimating each individual parameter separately it should be possible to achieve independence of estimated parameters, in the sense of vanishing off-diagonal elements of the covariance matrix, so that imperfect knowledge of one of them does not deteriorate the precision of estimating the others.
If these three conditions are satisfied, the optimal scheme for any of the parameters individually is no more powerful than the scheme in which they are all estimated together and
we say that the channel parameters to be estimated are \emph{compatible}.

\begin{figure}[t]
\includegraphics[width=0.9 \columnwidth]{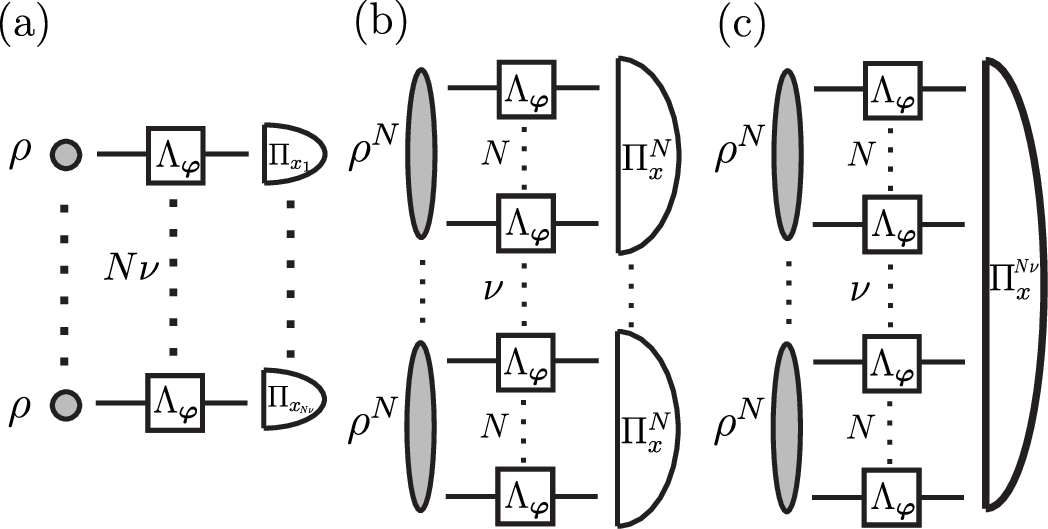}
\caption{Three scenarios of utilizing $n=N \nu$ quantum probes in metrology: (a) ``classical'' scheme, where both input probes and measurements are uncorrelated, resulting in $n$ independently  and identically distributed random variables $x_i$, (b) entangled probe scheme, where $\rho^N$ represents states on a Hilbert space $\mathcal{H}^{\otimes N}$, where $\mathcal{H}$ is the space upon which the channel $\Lambda_{\gr{\varphi}}$ acts, and measurements occur on the level of individual $\rho^N$ (c) collective measurement scheme, where input probes may be arbitrarily entangled and collective measurements over arbitrarily many $\rho^N$ are allowed.}
\label{fig:metrologycorrleations}
\end{figure}

In the above, we have not yet discussed the role of entanglement in the state preparation and measurement stages. We represent three relevant scenarios relating to this in Fig.~\ref{fig:metrologycorrleations}, which can be regarded as more detailed illustrations of possible estimation schemes in Fig.~\ref{fig:metrology}, for example by letting the input state of Fig.~\ref{fig:metrology} (a) be  $\rho^{\otimes N\nu}$ and the channel $\Lambda^{\otimes N\nu}$, we get the same picture as Fig.~\ref{fig:metrologycorrleations} (a). In all of them we have the same number $n$ of channel applications, but in (b) and (c) subdivide the states into $\nu$ identical and independent blocks of $N$ arbitrarily entangled systems. 
In single parameter metrology, only Fig.~\ref{fig:metrologycorrleations} (a) and (b) are relevant; that is, scheme (c) holds no advantage over scheme (b). It is then known that for the QFI CR bound to be saturable \cite{Jarzyna2013} it is necessary to have many experimental repetitions, i.e. the bound is saturated as $\nu\to\infty$. For estimating single parameter unitary operations, scheme (b) allows the so-called Heisenberg limit of $\frac{1}{\nu N^2}$ scaling of variance to be attained, whereas (a) represents a shot-noise limited experiment with scaling $\frac{1}{\nu N}$.

In multiparameter metrology, scenario (c) gains relevance, as allowing for collective measurement potentially provides an advantage. It remains important for our purposes that $\nu$ be large, as we use a result from the theory of quantum local asymptotic normality, which relies upon measurement of a large collection of identical and independent states. We discuss this further in the following.

\section{Multiparameter Cram{\'e}r-Rao bounds}
\label{sec:crbounds}
In this section we review the main tools of multiparameter quantum metrology based on variants of CR bounds that are used further on in this paper. In particular we stress the difference between single and multiparameter cases as well discussing reasons why metrological incompatibility may appear in different settings.
\subsection{Classical Multiparameter Cram{\'e}r-Rao bound}
\label{sec:classicalCRbound}
First we shall consider a classical multiparameter estimation scheme. The central objects here are probability distributions $p(\gr{x}|\gr{\varphi})$ of data $\gr{x}$ dependent upon the parameters. This can be thought of as a quantum estimation problem where we've fixed a measurement $\lbrace \Pi_{\gr{x}}\rbrace$ and state, thereby obtaining $p(\gr{x}|\gr{\varphi})=\tr{\rho_{\gr\varphi} \Pi_{\gr{x}}}$. We can define the Fisher information (FI) matrix for $m$ parameters as the $m \times m$ matrix with entries given by
\begin{equation}
F_{ij}(\boldsymbol\varphi)= \sum_{\gr{x}} p(\gr{x}|\boldsymbol\varphi) \left(\frac{\partial\text{ln} p(\gr{x}|\boldsymbol\varphi)}{\partial \varphi_i}\right)\left(\frac{\partial\text{ln} p(\gr{x}|\boldsymbol\varphi)}{\partial \varphi_j}\right).
\end{equation}
Crucially, this matrix allows us to define the multiparameter CR bound:
\begin{equation}
\label{eq:classicalCR}
\text{Cov}(\gr{\tilde{\varphi}})\geq F^{-1}(\boldsymbol\varphi),
\end{equation}
where $\text{Cov}(\gr{\tilde{\varphi}})$ refers to the covariance matrix for  a locally unbiased estimator $\gr{\tilde{\varphi}}(\gr{x})$, $\text{Cov}(\gr{\tilde{\varphi}})_{ij} = \langle(\tilde{\varphi}_i - \varphi_i)(\tilde{\varphi}_j - \varphi_j) \rangle$ and $\langle \cdot \rangle$ represents the average with respect to the probability distribution $p(\gr{x}|\gr{\varphi})$. The above inequality should be understood as a matrix inequality. In general, we can write $\tr [G\text{\,Cov}(\gr{\tilde{\varphi}})]\geq \tr (GF^{-1}(\boldsymbol\varphi))$ where $G$ is some positive cost matrix, which allows us to asymmetrically prioritise the uncertainty cost of different parameters. As in the single parameter case, the bound is saturable in the limit of an infinite number of repetitions of an experiment using the maximum likelihood estimator \cite{Kay1993}.

The first substantial difference of multiparameter metrology from the single parameter case can already be discussed  at the classical level.
Assuming we've already chosen a probe state and a measurement, it may happen that the resulting FI matrix is non-diagonal. This means that the estimators for the parameters will not be independent. Considering now the separate schemes of Fig.~\ref{fig:metrology} (b) and assuming all parameters except the $i$-th one  are perfectly known, the single parameter CR bound implies that the uncertainty of estimating
the $i$-th parameter is lower bounded by $\t{Var}(\tilde{\varphi}) \geq 1/F_{ii}$. On the other hand in the simultaneous scenario of Fig.~\ref{fig:metrology}b according to  \eqref{eq:classicalCR} we have $\t{Var}(\tilde{\varphi}) \geq (F^{-1})_{ii}$. From  basic algebra of  positive-definite matrices, we have that
$(F^{-1})_{ii} \geq 1/F_{ii}$, with equality holding only in the case when all off-diagonal elements $F_{ij}=0$, $j\neq i$.
Since asymptotically the CR bound is saturable, it implies that equal performance between the simultaneous and $p$ separate schemes in the limit of a large number of experiment repetitions can only hold if $F$ is a diagonal matrix, and hence there are no statistical correlations between the estimators \cite{Cox}. Otherwise condition (iii) for parameter compatibility is violated.

Clearly, for any real positive definite matrix one can perform an orthogonal rotation to a new basis in which the matrix is diagonal. This simply means that there are always linear combinations of the parameters for which the diagonality conditions hold. Often, however, the choice of the parameters we are interested in arise as a result of physical considerations and in this sense there is a preferred basis in which the question of parameter compatibility has clear physical implications.

\subsection{Quantum Fisher Information Cram{\'e}r-Rao bound}

While the fundamental objects for calculating the classical FI are probability distributions of the data conditioned on the parameters to be estimated, the fundamental objects in the quantum problem are the density matrices $\rho_{\gr{\varphi}}$ dependent on these parameters. Note that here we assume that a probe state has already been selected and subjected to evolution and hence for the time being we ignore the issue of optimization over input probes.

In the quantum scenario we therefore face an additional challenge of determining the optimal measurement for extracting most of the information on the parameters of interest from the quantum states. In the single parameter case the
situation is relatively simple. Maximization of the classial FI over all quantum measurements yields the quantity referred to as the QFI
which can be calculated using the following formula:
\begin{equation}
F_Q(\varphi)= \text{tr}(\rho_\varphi L^2),
\end{equation}
where $L$ is a Hermitian matrix, the so-called symmetric logarithmic derivative (SLD), defined implicitly by \
$\frac{1}{2}(L \rho_\varphi + \rho_\varphi L)= \partial_\varphi \rho_\varphi$, where for simplicity of notation we do not explicitly write
the dependence of $L$ on $\varphi$.  Moreover, one can always choose the projective measurement in the eigenbasis of the SLD which
 yields FI equal to the QFI. Hence, the QFI determines the ultimate achievable precision of estimating the parameter on
density matrices $\rho_\varphi$ in the asymptotic limit of an infinite number of experiment repetitions.
Moreover, the fact that the QFI is additive on tensor product density matrices, in particular ${F_Q}(\rho_\varphi^{\otimes N})=N {F_Q}(\rho_\varphi)$,
and achievable via individual measurements, implies that there is no asymptotic gain in performing collective measurements over individual ones, hence scenarios (b) and (c) in Fig.~\eqref{fig:metrologycorrleations} are equivalent in the single parameter estimation case.

We now move on to a multiparameter scenario. A direct generalization of single parameter CR bound leads to the multiparameter QFI CR bound \cite{Helstrom, Holevo} that reads:
\begin{equation}
\t{Cov}(\tilde{\gr{\varphi}}) \geq  {F_Q}(\gr{\varphi})^{-1}, \quad
 {F_Q}_{ij}(\boldsymbol\varphi)=\frac{1}{2}\text{tr}(\rho_{\boldsymbol{\varphi}} \lbrace L_i,L_j \rbrace),
\end{equation}
where the braces refer to the anticommutator, whereas $L_i$ is the SLD related to parameter $i$, defined analogously to the single parameter case as $\frac{1}{2}(L_{i} \rho_{\gr{\varphi}} + \rho_{\gr{\varphi}} L_{i})= \partial_{\varphi_i}\rho_{\gr{\varphi}}$.
As a result, given any cost matrix $G$, the estimation cost is bounded by,
\begin{equation}
\label{eq:multiCRbound}
\tr[G \cdot \t{Cov}(\tilde{\gr{\varphi}})] \geq \tr(G {F_Q}^{-1}).
\end{equation}
Unlike in the single parameter case the above bound is not always saturable. The intuitive reason for this is incompatibility of the optimal measurements for different parameters.
Under what conditions may we nevertheless hope to saturate the bound?
Given that the optimal measurement for a given parameter is formed from projectors corresponding to the eigenbasis of the SLD, we may immediately identify that if $[L_i,L_j]=0$ then there is a single eigenbasis for both SLDs and thus a common measurement optimal from the point of view of extracting information on $\varphi_i$ as well as $\varphi_j$. However, this is only a sufficient but not a necessary condition.
We discuss a necessary and sufficient condition in Sec.~\ref{sec:saturability}, but in preparation for this, we need
to introduce a more powerful version of the multiparameter CR bound.

\subsection{Holevo Cram{\'e}r-Rao Bound}
\label{sec:holevoCRbound}
The problem with saturability of the multiparameter QFI CR bound was realized early in the development of quantum estimation theory by Holevo \cite{Holevo}. He proposed a stronger multiparameter bound which we refer to as the Holevo CR bound. Its original formulation is not very explicit and therefore we prefer to use its equivalent formulation put forward in \cite{Nagaoka1989}.
Given a cost matrix $G$ the achievable estimation uncertainty is lower bounded by
\begin{equation}
\label{eq:boundholevo}
\tr[G \cdot \t{Cov}(\tilde{\gr{\varphi}})] \geq \min_{\{X_i\}}\left\{\tr(G \cdot \t{Re} V) +\|\sqrt{G} \cdot \t{Im}V\cdot\sqrt{G}\|_1 \right\},
\end{equation}
where $\| \cdot\|_1$ is the trace norm, $V_{ij} = \tr(X_i X_j \rho_{\gr{\varphi}})$,
 and the minimization is performed over Hermitian matrices $X_i$, satisfying $\frac{1}{2}\tr(\{X_i,L_j\}
 \rho_{\gr{\varphi}}) = \delta_{ij}$, where $L_i$ are SLDs as defined before.
 The last constraint plays the role of the local unbiasedness condition.

This bound is indeed stronger than the QFI CR bound which may be appreciated by rewriting the r.h.s. of the QFI
bound, Eq.~\eqref{eq:multiCRbound}, in the following form \cite{Nagaoka1989}:
\begin{equation}
\label{eq:boundQFIalternative}
\tr(G  {F_Q}^{-1}) = \min_{\{X_i\}}\tr(G \cdot \t{Re} V),
\end{equation}
with the same constraints on the $X_i$ matrices as in the definition of the Holevo CR bound. Clearly, since the second term in Eq.~\eqref{eq:boundholevo} is positive, it implies that the QFI bound is in general weaker.
As the above formula for the QFI CR bound is not widely recognized, for the sake of completeness and anticipating
further discussion of the saturability issue, we provide a proof of it below.

Let us write the solution to the minimization problem of the r.h.s of Eq.~\eqref{eq:boundQFIalternative}
explicitly using the Lagrange multiplier method. Introducing Lagrange multipliers $\lambda_{ij}$ we need to minimize
\begin{equation}
\frac{1}{2}\sum_{ij} G_{ij} \tr(\rho_{\gr{\varphi}}\{X_i, X_j\})  -
\lambda_{ij}[\delta_{ij}- \frac{1}{2}\tr(\rho_{\gr{\varphi}}\{X_i, L_j\})]
\end{equation}
over Hermitian $X_i$. Each $n$-dimensional Hermitian matrix $X_i$ may be parametrized by $n^2$ real parameters. Taking the derivatives over each of these produces a set of matrix equations,
\begin{equation}
\label{eq:lagrange}
\forall_i \sum_j G_{ij}\{\rho_{\gr{\varphi}},X_j\} - \lambda_{ij} \{\rho_{\gr{\varphi}}, L_j\} = 0.
\end{equation}
Taking
\begin{equation}
X_i = \sum_j (G^{-1} \Lambda)_{ij} L_j.
\end{equation}
where by $\Lambda$ we denote the matrix of Lagrange multipliers $(\Lambda)_{ij} = \lambda_{ij}$ it is clear that Eq.~\eqref{eq:lagrange}
is satisfied. Moreover, the constraint condition $\frac{1}{2}\tr(\{X_i,L_j\}
 \rho_{\gr{\varphi}}) = \delta_{ij}$ reads:
\begin{equation}
\frac{1}{2} (G^{-1} \Lambda)_{ik} \tr(\{L_k,L_j\}
 \rho_{\gr{\varphi}}) = \delta_{ij}.
\end{equation}
This implies that the Lagrange multiplier matrix must be chosen so that:
\begin{equation}
G^{-1} \Lambda  {F_Q} = \openone.
\end{equation}
As a result the solution to the minimization problem reads
\begin{equation}
X_i = \sum_{j}({F_Q}^{-1})_{ij} L_j
\end{equation}
and utilizing the fact that QFI matrix is symmetric we get
\begin{equation}
\tr(G \cdot \t{Re} V) = \tr(G {F_Q}^{-1} {F_Q} F_Q^{-1}) = \tr(G F_Q^{-1}),
\end{equation}
which ends the proof.

Even though the Holevo CR bound is tighter than the QFI one, it is still not always saturable with separable measurements.
However, it is saturable for Gaussian state shift models where the
parameters are encoded in the first moment displacements \cite{Holevo}. Even more interestingly,
thanks to the theory of quantum local asymptotic normality (QLAN)  \cite{Hayashi,Kahn,Yamagata} which
 asymptotically maps any quantum estimation problem performed on a large number of copies of a quantum state
 to a corresponding Gaussian shift model, the Holevo CR bound is asymptotically achievable
  in this case as well. Since the mapping does not respect separation into single copy subsystems,
  collective measurement may in general be required to saturate the Holevo CR bound.
  Hence, for all schemes depicted in Fig.~\ref{fig:metrologycorrleations} (c)
  the Holevo CR bound provides an ultimate asymptotically saturable multiparameter CR bound.

\section{Multiparameter compatibility}
\label{sec:saturability}
\subsection{Saturability of multiparameter Quantum Fisher Information Cram{\'e}r-Rao bound}
As we mentioned before, if the SLDs $L_i$ corresponding to the different parameters commute, there is no additional difficulty in extracting optimal
information from a state on all parameters simultaneously. If they do not commute, however, this does not immediately imply that
it is impossible to simultaneously extract information on  all parameters with precision matching that of the separate scenario for each.

A weaker condition has appeared in a number of papers \cite{Matsumoto, Meshloss,Mesh, Genoni,Suzuki}
which states that the multiparameter QFI CR bound can be saturated provided
\begin{equation}
\label{Commutation}
\tr(\rho_{\gr{\varphi}} [L_{i},L_{j}])=0,
\end{equation}
where not the commutator itself but only its expectation value on the probe state is required to vanish.
Henceforth we shall refer to this as the \textit{commutation condition}.
This condition was first identified as necessary and sufficient by Matsumoto \cite{Matsumoto} for the case  when $\rho_{\gr{\varphi}}$ is a pure state, upon which the criterion is equivalent to the existence of \textit{some} pair of SLDs which commute, given that SLDs are not unique on pure states. It is then possible to find an optimal measurement as the common eigenbasis of these SLDs. This implies that for unitary evolution on pure states, satisfaction of the commutation condition coincides with the existence of commuting Hamiltonians which \textit{could} have generated the evolution on the given probe.

For mixed states, this criterion has been identified in a comprehensive characterisation of the behaviour of the Holevo bound for two-parameter estimation on separable qubits \cite{Suzuki}. Elsewhere, it has been used in more general settings but without a readily available proof which we are aware of and has met some small inconsistencies in its usage, being variously identified as sufficient \cite{Genoni} or necessary and sufficient \cite{Meshloss} in different papers. To clear up this confusion we present a derivation of this criterion,
which to the best of our knowledge has not been provided before in such a simple, direct and general manner.

First of all, we consider a scenario where estimation is performed on multiple independent copies of the output state $\rho_{\gr{\varphi}}$
and allow for collective measurements as in Fig.~\ref{fig:metrologycorrleations} (c). We know already from the discussion in Sec.~\ref{sec:holevoCRbound} that in this case the Holevo CR bound is asymptotically achievable thanks to QLAN theory.
Hence, to prove asymptotic saturability of the  multiparameter QFI CR bound it is enough to prove that it is  equivalent to the Holevo CR bound
if and only if the commutation condition \eqref{Commutation} holds.

\paragraph*{Proof}
For the sake of the proof we assume that both the cost matrix $G$ and QFI matrix $F_Q$ are strictly positive. These are natural assumptions since otherwise if some eigenvalues of $G$ were zero,  uncertainty in some parameter combinations would not be penalized whereas if some
eigenvalues of $F_Q$ were zero, it would be impossible to estimate some of the parameters with finite precision.

Let us first prove  sufficiency of \eqref{Commutation} and assume that $\tr([L_{i},L_{j}] \rho_{\gr{\varphi}})=0$.
We have seen that when calculating the minimum in the formula for the QFI bound using \eqref{eq:boundQFIalternative}
we have found that the optimal $X_i = \sum_{j}({F_Q}^{-1})_{ij} L_j$ are linear combinations of $L_i$.
Since $\tr([L_{i},L_{j}] \rho_{\gr{\varphi}})=0$ for all $i,j$ it implies that the the same holds for all
their linear combinations and hence $\tr([X_{i},X_{j}] \rho_{\gr{\varphi}}) = 0$ for all $i,j$. This, however, implies that the same set of $X_i$ minimizes the formula for the Holevo bound as it makes the second term in \eqref{eq:boundholevo} equal to zero.

To prove the necessity we assume that the Holevo bound coincides with the QFI bound and hence for the $X_i$ that minimize
both \eqref{eq:boundholevo} and \eqref{eq:boundQFIalternative}  the second term in \eqref{eq:boundholevo} must be equal to zero.
Since  $G$ is strictly positive, this implies
that the matrix $\t{Im}V$ must be zero and hence $\tr([X_{i},X_{j}] \rho_{\gr{\varphi}}) = 0$ for all $i,j$.
On the other hand, we know that the $X_i$ minimizing \eqref{eq:boundQFIalternative} have the form
$X_i = \sum_{j}({F_Q}^{-1})_{ij} L_j$. Inverting this formula we get $L_i = \sum_{j}({F_Q})_{ij} X_j$ and hence
$\tr([L_{i},L_{j}] \rho_{\gr{\varphi}})=0$ for all $i,j$  $\blacksquare$.

It's worth stressing the different implications of the commutation condition on pure states and on mixed states. In the case of pure states,
as already mentioned above, the commutation relation implies that there is an individual measurement that allows saturation of the QFI CR bound as in Fig.~\ref{fig:metrologycorrleations} (b).
On the other hand, for mixed states, collective measurements on multiple copies may be necessary in general to achieve the bound as in Fig.~\ref{fig:metrologycorrleations} (c).
This is due to the fact that the Holevo CR bound is guaranteed to be saturable provided one takes the asymptotic limit of many independent copies of a state, while the correspondence to Gaussian states via QLAN theory implicitly does not invoke limitations
on the allowed set of measurements.

\subsection{Conditions for multiparameter compatibility}

Combining the commutation condition with the parameter independence condition discussed in Sec.~\ref{sec:classicalCRbound} which requires off-diagonal QFI matrix entries to be zero, we  arrive at a necessary requirement for multiparameter compatibility which reads
\begin{equation}
\label{eq:compatibilty}
\forall_{i \neq j} \tr(L_{i}L_{j} \rho_{\gr{\varphi}})=0.
\end{equation}
Plugging in an explicit form for the SLDs
\begin{align}
L_{i}=2\sum_{m,n} \frac{\langle\psi_m|(\partial_{\varphi_i} \rho_{\gr{\varphi}}) |\psi_n\rangle}{p_m+p_n}|\psi_m\rangle\langle\psi_n|,
\end{align}
where $p_{m,n}$ and $|\psi_{m,n}\rangle$ are the eigenvalues and eigenvectors of the state $\rho_{\gr{\varphi}}=\sum_k p_k |\psi_k\rangle\langle\psi_k|$ from which parameters are to be estimated.
The compatibility condition \eqref{eq:compatibilty} can now be written as:
\begin{equation}
\label{eq:compatibilityexplicit}
\forall_{i \neq j}\sum_{m,n}\frac{p_m}{(p_m+p_n)^2}\langle\psi_m|\partial_{\varphi_i} \rho_{\gr{\varphi}}
|\psi_n\rangle \langle\psi_n|\partial_{\varphi_j} \rho_{\gr{\varphi}} |\psi_m\rangle = 0.
\end{equation}

On top of this we must not forget the final condition which demands the existence of a single probe state that provides maximum
QFIs for all the parameters.

In summary, we may decompose the demands of simultaneous estimation into several layers of stringency. The first is the existence of a single probe state yielding maximum possible values of QFIs  for all parameters of interest. Second is the requirement of the
existence of compatible measurements on the output states which ensures the saturability of the QFI CR bound and the last one
is the requirement that the QFI matrix is diagonal which enables independent estimation of the parameters.

If all these conditions hold, the optimal metrological strategy will not depend on the choice of the cost matrix $G$ and
 the ultimate  bounds on estimation precision are found in the same way as in the case of single parameter estimation.

\section{Unitary parameter estimation}
\label{sec:unitary}
Let us first treat the case of multiple unitary parameter estimation, which has been considered in a number of papers \cite{Bagan2001d, Chiribella2004, Kolenderski2008,Genoni,Ballester, two-step, Fujiwara,Multiphase, Berry2015, Baumgratz2016}
and ask under what conditions we can have multiparameter compatibility.
We consider unitary evolution acting on the input probe state to be of the form
\begin{equation}
U_{\gr{\varphi}} = e^{\mathrm{i} \sum_k H_k \varphi_k}.
\end{equation}
Thanks to convexity of the QFI we can always assume the input state to be pure $\rho=\ket{\psi}\bra{\psi}$.
Since the evolution is unitary, the output state will be pure as well $\ket{\psi_{\gr{\varphi}}} = U_{\gr{\varphi}} \ket{\psi}$.
For pure states the SLDs can be explicitly written as:
\begin{equation}
L_i = 2(\ket{\psi^{(i)}_{\gr{\varphi}}}\bra{\psi_{\gr{\varphi}}}+ \ket{\psi_{\gr{\varphi}}}\bra{ \psi^{(i)}_{\gr{\varphi}}})
\end{equation}
where $\ket{\psi^{(i)}_{\gr{\varphi}}} = \partial_{\varphi_i} \ket{\psi_{\gr{\varphi}}}$.
For the moment, for the sake of clarity we consider estimation performed around the point where all $\varphi_k=0$.
In this case
\begin{equation}
L_i = 2 \mathrm{i} ( H_i \ket{\psi_{\gr{\varphi}}}\bra{\psi_{\gr{\varphi}}} -  \ket{\psi_{\gr{\varphi}}}\bra{ \psi_{\gr{\varphi}}}H_i).
\end{equation}
As a result, the compatibility condition \eqref{eq:compatibilty} takes the form
\begin{equation}
\label{eq:compatibiiltyconditionhamiltonians}
\forall_{i \neq j} \bra{\psi} (\langle H_i \rangle - H_i)  (\langle H_j \rangle - H_j) \ket{\psi} = 0,
\end{equation}
where $\langle H_i \rangle = \bra{\psi} H_i \ket{\psi}$.
Additionally, apart from fulfilling the above orthogonality conditions we must make sure that the single input probe
yields optimal QFI with respect to all parameters. The QFI for the $i$ parameter is simply proportional to the variance
of $H_i$:
\begin{equation}
(F_{Q})_{ii} = \bra{\psi} (\langle H_i \rangle - H_i)^2 \ket{\psi}
\end{equation}
and is uniquely maximized by a probe state which is an equally weighted superposition of eigenstates $\ket{-}_i$, $\ket{+}_i$ of $H_i$ corresponding to the minimal and the maximal eigenvalues $\lambda^-_{i}$, $\lambda_{i}^+$ respectively  \cite{GioMac}
\begin{equation}
\label{eq:optimalstatemultiunitary}
\ket{\psi} = \frac{1}{\sqrt{2}}(\ket{-}_i + \ket{+}_i).
\end{equation}
    The above form of $\ket{\psi}$ should be valid irrespectively of index $i$. Clearly, we have freedom to adjust the relative phases in the above expression, but we can also assume that they are incorporated in the definition of the eigenstates themselves.
Without losing generality, let us shift the Hamiltonians $H_i\to H_i - \frac{\lambda_i^++\lambda_i^-}{2}\mathbb{1}$ so that $\lambda_i^{-} = - \lambda_i^+ = - \lambda_i$ and hence
$\langle H_i \rangle = 0$ on the optimal probe state. Plugging the form of the optimal state \eqref{eq:optimalstatemultiunitary}
into \eqref{eq:compatibiiltyconditionhamiltonians} we get
\begin{equation}
\forall_{i \neq j} (\bra{+}_i - \bra{-}_i)(\ket{+}_j - \ket{-}_j) = 0.
\end{equation}
After some basic algebra this implies that the extremal eigenvectors of $H_i$ must necessarily be of the form:
\begin{eqnarray}
\label{eq:unitaryeigenvectorscompatibility}
\ket{+}_i & =& \frac{1}{\sqrt{2}}(\ket{\psi} +  \ket{\xi_i}),\\
\ket{-}_i & =& \frac{1}{\sqrt{2}}(\ket{\psi} -  \ket{\xi_i}),
\end{eqnarray}
where $\braket{\xi_i}{\xi_j} = \delta_{ij}$ and all $\ket{\xi_i}$ are orthogonal to $\ket{\psi}$.
The above formulas express the most general requirements on the eigenvectors of the generating Hamiltonians for compatible metrology to be achievable in this evolution model.
As will be shown in the example below these are rather stringent
conditions. One might object that e.g. in the case where all generators $H_i$ are equal there should be no difficulty in estimating
simultaneously multiple parameters since the optimal input probe state and the optimal measurements are identical for all $\varphi_i$.
Note however that such a model provides us only with the information on the total accumulated phase $\sum_i \varphi_i$
and therefore the statistical independence condition is not satisfied, and even worse, the QFI matrix is degenerate.

To end this general discussion, let us go back to the more general case of $\varphi_i \neq 0$.
In this case all the above discussion is valid up to replacement of all $H_i$ operators appearing in formulas from Eq.~\eqref{eq:compatibiiltyconditionhamiltonians} onwards with
$H_i^S=U^\dagger_{\boldsymbol{\varphi}}\mathcal{S}_i[H_i e^{\mathrm{i} \sum_k H_k \varphi_k}]$, where $\mathcal{S}_i$ represents
a symmetrization operation which acts when encountering any product of $H_i$ with other operators $H_{k \neq i}$ that do not commute with it. It performs
a normalized symmetrization of this product, so e.g. $\mathcal{S}_1[H_1 H_2^2] = \frac{1}{3}(H_1 H_2^2 + H_2 H_1 H_2 + H_2^2 H_1)$.
The above considerations may also be easily adapted to the case where different parameter unitaries act sequentially i.e.
$U_{\gr{\varphi}}= \Pi_k e^{\mathrm{i} H_k \varphi_k}$, by replacing $H_i^S$  with
$(\Pi_{k=1}^{i-1}e^{\mathrm{i} H_k \varphi_k}) H_i (\Pi_{k=i}^{p}e^{\mathrm{i} H_k \varphi_k})$.

\subsection{Two-parameter estimation of a spin rotation}
Let us consider a spin $j$ particle, with associated angular momentum operator $\vec{S}=(S_x,S_y,S_z)$
and consider unitary two-parameter evolution of the form:
\begin{equation}
U_{\varphi_1,\varphi_2} = e^{\mathrm{i} \varphi_1 \vec{n}_1 \cdot \vec{S} +\mathrm{i} \varphi_2 \vec{n}_2 \cdot \vec{S}},
\end{equation}
where the $H_i$ generating the unitary transformation now correspond to different directions of the spin operators $H_i= \vec{n}_i \cdot \vec{S}$.
For simplicity we focus on estimation around $\varphi_1=\varphi_2=0$ point, though the discussion remains qualitatively equivalent
 when $\varphi_i \neq 0$.
Let $\ket{m}_{\vec{n}}$, $m \in {-j,\dots,j}$ denote the basis constructed from eigenvectors of the $\vec{n}\cdot \vec{S}$ operator with
projection value $m$. According to previous discussion the optimal state needs to have the form
\begin{equation}
\ket{\psi} = \frac{1}{\sqrt{2}}(\ket{{-j}}_{\vec{n}_1}+ \ket{{+j}}_{\vec{n}_1}) = \frac{1}{\sqrt{2}}(\ket{{-j}}_{\vec{n}_2}+ \ket{{+j}}_{\vec{n}_2}),
\end{equation}
and clearly $\langle H_i \rangle = 0$.
Let $\alpha$ be the angle between directions $\vec{n}_1$ and $\vec{n}_2$.
Using standard theory of angular momentum we may expand states $\ket{\pm j}_{\vec{n}_2}$ in
the basis $\ket{m}_{\vec{n}_1}$ as follows
\begin{align}
\ket{{+j}}_{\vec{n}_2} &= \sum_{m=-j}^j {\tbinom{2j}{j+m}}^{\frac{1}{2}} \sin^{j+m}\tfrac{\alpha}{2} \cos^{j-m}\tfrac{\alpha}{2} \ket{m}_{\vec{n}_1}, \\
\ket{{-j}}_{\vec{n}_2} & = \sum_{m=-j}^j (-1)^{j-m}\tbinom{2j}{j+m}^{\frac{1}{2}} \cos^{j+m}\tfrac{\alpha}{2} \sin^{j-m}\tfrac{\alpha}{2} \ket{m}_{\vec{n}_1},
\end{align}
where we have neglected any possible relative phases that might appear in the above decomposition as they are irrelevant in the following.
Rewriting the formula for $\ket{{+j}}_{\vec{n}_2}$ as
\begin{equation}
\ket{{+j}}_{\vec{n}_2} = \cos^{2j}\tfrac{\alpha}{2}\ket{-j}_{\vec{n}_1} + \sin^{2j}\tfrac{\alpha}{2}\ket{j}_{\vec{n}_1} + \sum_{m=-j+1}^{j-1} \dots
\end{equation}
and comparing it with the compatibility conditions \eqref{eq:unitaryeigenvectorscompatibility} we
see that the only possibility of satisfying them is to take $\alpha=\pi/2$ and $j=1$ in which case we obtain
\begin{eqnarray}
\ket{{+1}}_{\vec{n}_2} &=& \frac{1}{2}(\ket{-1}_{\vec{n}_1} + \ket{1}_{\vec{n}_1}) + \frac{1}{\sqrt{2}}\ket{0}_{\vec{n}_1}, \\
\ket{{-1}}_{\vec{n}_2} &=& \frac{1}{2}(\ket{-1}_{\vec{n}_1} + \ket{1}_{\vec{n}_1}) - \frac{1}{\sqrt{2}}\ket{0}_{\vec{n}_1},
\end{eqnarray}
resulting in estimation precision $\Delta^2 \varphi_1 = \Delta^2\varphi_2 = 1/4$.
With this example it is clear how restrictive the multiparameter compatibility conditions in metrology are.
The fact that for spin $j=1/2$ there is no possibility for satisfying compatibiliy conditons is clear from
\eqref{eq:unitaryeigenvectorscompatibility} as at least three dimensional space is required to have three orthogonal states
$\ket{\psi}$, $\ket{\xi_1}$, $\ket{\xi_2}$. It is, however, nontrivial that the only
case where multiparameter compatibility can be satisfied is $j=1$ for rotations around two perpendicular axes.
Given that we are working in the pure state case, it is always possible to find a measurement on a single spin that
achieves the quantum CRB. The following projection measurement suffices,
\begin{align*}
\Pi_1&=\frac{1}{2}(\ket{{+1}}_{\vec{n}_1}+ \ket{{-1}}_{\vec{n}_1})(\bra{{+1}}_{\vec{n}_1}+ \bra{{-1}}_{\vec{n}_1}) \,,\\
\Pi_2&=\frac{1}{2}(\ket{{+1}}_{\vec{n}_1}- \ket{{-1}}_{\vec{n}_1})(\bra{{+1}}_{\vec{n}_1}- \bra{{-1}}_{\vec{n}_1})\,,\\
\Pi_3&= \mathbb{1}-\Pi_1-\Pi_2.
\end{align*}
From the above discussion it is also clear that there is no possibility to estimate three different rotation directions in a compatible way since
the only promising case $j=1$ corresponds to a three dimensional space whereas compatibility of three different rotation parameters
require at least a four dimensional space according to \eqref{eq:unitaryeigenvectorscompatibility}.

The results presented above can be immediately applied to the case when $N$ qubits experience independent two parameter rotations
according to the following unitary
\begin{equation}
U_{\varphi_1,\varphi_2} = \left(e^{\frac{\mathrm{i}}{2} (\varphi_1 \vec{n}_1 \cdot \vec{\sigma} +\varphi_2 \vec{n}_2 \cdot \vec{\sigma})} \right)^{\otimes N},
\end{equation}
as in this case the optimal input probe state lives in the fully symmetric subspace which is isomorphic to spin $j=N/2$ space.
It is therefore clear that while for a single qubit ($N=1$) undergoing simultaneous rotation around two axes, the compatibility
conditions cannot be satisfied, they can be achieved when considering $N=2$ case and an appropriately chosen entangled input; essentially entanglement takes us from a highly incompatible case to full compatibility with Heisenberg scaling in two parameters at once (but only for $N=2$).
This fact can be confirmed by inspecting results presented in \cite{Genoni}, where the sum of variances of two angles
of rotations was minimized, and noticing that only in the case of $N=2$, the obtained result indeed corresponds to the optimal separate scenario.
For higher dimensional $N$ the Heisenberg bound is no longer achievable in both parameters. If we choose GHZ-type states, then we can achieve Heisenberg $1/N^2$ scaling in one parameter, but classical $1/N$ scaling in the other. Other states can achieve different trade-offs; for even-$N$ qubit Dicke states with $\frac{N}{2}$ excitations (in the direction mutually orthogonal to $\vec{n}_1$ and $\vec{n}_2$), both parameters have a Fisher information of $\frac{N^2}{2}+N$, which asymptotically retains quadratic scaling but with a $1/2$ prefactor.

\section{Hybrid unitary + non-unitary parameter estimation}
\label{sec:nonunitary}
In the previous section we have seen that the compatibility conditions in the case of multiple unitary parameters
are very demanding and can be satisfied only in very special situations.
 In this section we focus on the case when
 one of the parameters $\varphi$ is unitary whereas the other one, which we denote by $\eta$ enters via
 a non-unitary part of the evolution as e.g.  a decoherence strength parameter.

 This scenario has been considered before in several models such as the estimation of loss and phase in an interferometer \cite{Meshloss},
  as well as the estimation of phase with collective \cite{Durkin} and independent \cite{Mesh} dephasing. Here we want to investigate the possibility  of satisfying the compatibility conditions in such situations.

Before considering specific schemes, let us first identify some general sufficient criteria for the compatibility condition as expressed
by Eq.~\eqref{eq:compatibilty} and ignore for the  moment the requirement for the existence of common optimal input probe state.
The explicit form of the compatibility conditon \eqref{eq:compatibilityexplicit} can be written as:
\begin{equation}
\label{eq:compatibilityphieta}
\sum_{m,n}\frac{p_m}{(p_m+p_n)^2}\bra{\psi_m} \partial_{\varphi} \rho_{\varphi\eta}
\ket{\psi_n} \bra{\psi_n}\partial_{\eta} \rho_{\varphi\eta} \ket{\psi_m} = 0,
\end{equation}
where $p_n$, $\ket{\psi_n}$ are eigenvalues and eigenvectors of $\rho_{\varphi\eta}$.

Let us assume that the decoherence parameter $\eta$ induces a ``classical'' evolution in the sense that
\begin{equation}
\partial_\eta\rho_{\varphi\eta}=\sum_k (\partial_\eta p_k) |\psi_i\rangle\langle\psi_i|
\end{equation}
so that only the eigenvalues of the density matrix depend on the parameter and the state remains diagonal in its initial eigenbasis.
This makes all off-diagonal terms $m \neq n$ in \eqref{eq:compatibilityphieta} zero.  However, since the second
parameter is unitary,
\begin{equation}
\bra{\psi_n} \partial_\varphi \rho_{\varphi\eta} \ket{\psi_n} = p_n \partial_\varphi \braket{\psi_n}{\psi_n} = 0
\end{equation}
and hence the diagonal terms are zero as well, guaranteeing the compatibility condition to hold.



There are more involved cases when the decoherence parameter $\eta$ influences not only the eigenvalues but the form of the eigenvectors of $\rho_{\varphi,\eta}$ as well. It might happen that even though individual terms in \eqref{eq:compatibilityphieta} are non-zero they sum up to zero in the end. Such situations need to be dealt with on a case by case basis.

\subsection{Estimation of phase and loss in an interferometer.}
\label{sec:lossphase}
\begin{figure}[t]
\includegraphics[width=0.9\columnwidth]{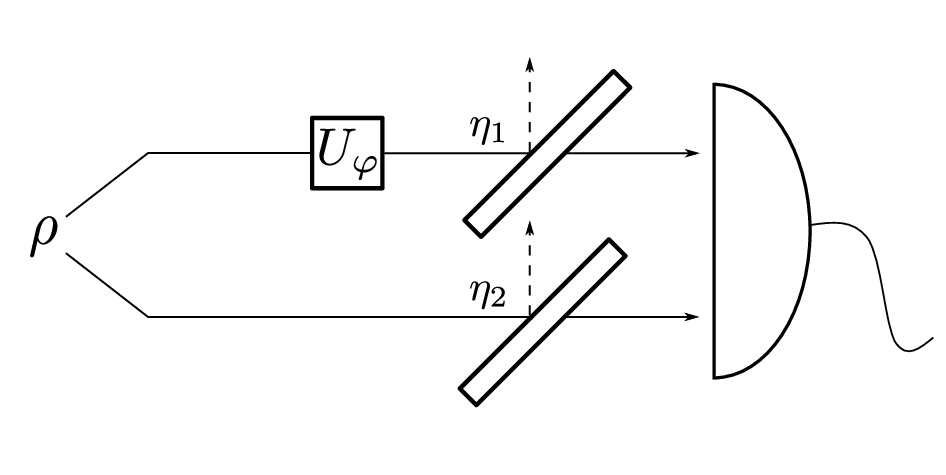}
\caption{A schematic of a general lossy interferometer with input state $\rho$. We model the losses by a beam-splitter. In \cite{Meshloss}, a scheme was considered with transmissivity $\eta_2=1$, leading to one arm containing both the loss and phase parameters. We balance the interferometer by choosing $\eta_1=\eta_2=\eta$.}
\label{interferometer}
\end{figure}

Consider an interferometer with equal loss in \textit{both} arms, as presented in Fig.~\ref{interferometer},
where the goal is to estimate both the relative phase delay $\varphi$ between the arms as well as the transmission coefficient $\eta$.
We choose for our input states to be fixed photon-number states, for which a general bipartite state is given by
\begin{equation}
|\psi\rangle=\sum_{k=0}^N \alpha_k |k,N-k\rangle.
\end{equation}
After passing through the interferometer the resultant state is
\begin{equation}
|\psi_{\varphi\eta}\rangle=\sum_{k=0}^N \sum_{l_2=0}^{N-k}\sum_{l_1=0}^k\alpha_k e^{i k \varphi} \sqrt{B^k_{l_1l_2}} |k,N-k\rangle \otimes |l_1,l_2\rangle,
\end{equation}
where the additional two modes represent photons lost from respectively the upper and the lower arm
and
\begin{equation}
B^k_{l_1l_2}= \binom{k}{l_1} \binom{N-k}{l_2} \eta^{N-l_1-l_2} (1-\eta)^{l_1+l_2}.
\end{equation}
On tracing out the auxiliary modes, we obtain a density matrix
\begin{equation}
\rho_{\varphi\eta}=\bigoplus_l \sum_{l_1} |\psi_{l_1,l-l_1}\rangle \langle \psi_{l_1,l-l_1}|,
\end{equation}
where different $l=l_1+l_2$ sectors represent different total number of photons lost while
\begin{equation}
|\psi_{l_1 l_2}\rangle= \sum_{k=l_1}^{N-l_2} \alpha_k e^{ik\varphi} \sqrt{B^k_{l_1l_2}} |k-l_1,N-k-l_2\rangle
\end{equation}
are subnormalized states corresponding to the situation of losing $l_1$ and $l_2$ photons in the upper and the lower arm respectively.
Note that states $\ket{\psi_{l_1,l-l_1}}$ living in a single $l$ sector are in general not orthogonal and hence should not be understood
as eigenvectors of  $\rho_{\varphi\eta}$.
Still, owing to the fact that
\begin{equation}
\partial_\eta B^k_{l_1l-l_1}=c_{N,l} B^k_{l_1l-l_1}, \quad c_{N,l} = \tfrac{N-l}{\eta}-\tfrac{l}{1-\eta}
\end{equation}
we eventually arrive at:
\begin{equation}
\partial_\eta\rho_{\varphi\eta}=
\bigoplus_l c_{N,l}\sum_{l_1} |\psi_{l_1,l-l_1}\rangle\langle\psi_{l_1,l-l_1}|,
\end{equation}
implying that upon differentiation the whole block corresponding to a fixed $l$ is multiplied by the\
  same constant factor. This means that only the eigenvalues of the density matrix are changed
  with variations of the parameter $\eta$ and hence we conclude that variations of $\eta$ induce the ``classical'' evolution. From the general considerations presented in the beginning of this section this implies that the compatibility criterion is satisfied.

More specifically, a brief calculation shows that the SLD $L_\eta$ decomposes into a weighted sum of projectors onto the blocks of the density matrix and thus an optimal measurement for loss is simply the set of projectors onto each block of constant $l$. The resultant Fisher information reads $\sum_L c_{Nl}^2 P(l|\eta)$, where most importantly
$P(l|\eta)=\tr \rho_{\varphi\eta} \Pi_l= \binom{N}{l}\eta^{N-l}(1-\eta)^l$ does not depend on the input state.
As a result we simply get a binomial distribution of total numbers of photons lost, and sampling this is the most informative thing we can do
 to learn $\eta$. The corresponding QFI reads $(F_Q)_{\eta\eta}=\frac{N}{\eta(1-\eta)}$.

Since $\partial_\varphi\rho_{\varphi\eta}$ does not mix blocks of different total photon-number (as phase shifts do not alter photon number), we find that $L_\varphi$ can be decomposed into the same blocks as $L_\eta$, and since $L_\eta$ simply acts as a multiple of the identity block-wise, they properly commute, not just under expectation value. Hence no collective measurements on multiple
copies of the quantum state are necessary to saturate the QFI CR bound, even though we are in the mixed state case.

Finally, we do not face the problem of determining a common optimal input probe. Since precision of estimating $\eta$ is state independent we
simply take the optimal state maximizing QFI for phase estimation \cite{Rafal,Jarzyna2013,Knysh2010}.
Taking the asymptotic analytical formula for optimal QFI in the limit of large $N$ \cite{Kolodynski2010, Knysh2010, Escher2011, Jan} and assuming $\eta < 1$ we summarize this section
by providing the achievable precision of compatible simultaneous phase and loss estimation:
$\Delta^2\varphi = \frac{1-\eta}{\eta N}$, $\Delta^2 \eta = \frac{\eta(1-\eta)}{N}$.


\subsection{Estimation of phase and dephasing}
Let us now consider $N$ qubits undergoing evolution composed of unitary phase combined with individual
dephasing processes. Each qubit is affected independently and the output $N$-qubit density matrix reads:
\begin{equation}
\rho_{\varphi\eta} = \Lambda_{\varphi\eta}^{\otimes N}(\rho), \label{Krauss}
\end{equation}
where
\begin{equation}
\Lambda_{\varphi\eta}(X) = U_\varphi\left(\sum_{i=0}^1 K_i X K_i^\dagger\right) U_\varphi^\dagger ,
\end{equation}
$U_\varphi =\exp(\mathrm{i}\varphi \sigma_z/2) $, while the two Kraus operators read $K_0=\sqrt{\frac{1+\eta}{2}}\mathbb{1}$ and $K_1=\sqrt{\frac{1-\eta}{2}}\sigma_z$.

In the case of $N=1$, any state on the equator of the Bloch sphere is known to be optimal both from the point of view of estimating phase
as well as the dephasing coefficient  \cite{Tesio}. Taking $\rho=\ket{+}\bra{+}$, with $|+\rangle=\frac{1}{\sqrt{2}}(|0\rangle+|1\rangle)$
we find the output state
\begin{equation}
\rho_{\varphi\eta} = \eta \ket{\varphi}\bra{\varphi} + (1-\eta)\openone/2,
\end{equation}
where $\ket{\varphi} = \frac{1}{\sqrt{2}}(|0\rangle+e^{\mathrm{i}\varphi}|1\rangle)$.
This is clearly the case where $\eta$ induces ``classical'' evolution, changing the eigenvalues without changing the eigenvectors and hence
the compatibility condition is immediately satisfied.

Still, as discussed in detail in   \cite{Mesh}, saturating the QFI CR bound in this case requires application of collective measurement on multiple copies of the state, unlike in the example of estimating phase and loss.
Discussion in \cite{Mesh} was restricted to probes being products of single qubit states. Here we want
to investigate the problem of simultaneous estimation in case of arbitrary entangled input states of $N$ qubits, since utilizing
entangled input probes is indispensable to reach the optimal phase estimation performance in the presence of dephasing \cite{Huelga, Orgikh2001, Escher2011}. It is known that the optimal input states are highly symmetric, exhibiting both permutational symmetry of the qubits, and also a parity symmetry under bit flips, i.e. they are invariant under $\sigma_x^{\otimes N}$ where $N$  refers to the number of qubits.
We will thus investigate the class of $N$-qubit states defined by these two kinds of symmetries.
This assumption is further justified by the fact the states optimized from the point of view of estimating the dephasing coefficient satisfy these symmetries as they are simply the product states $\ket{+}^{\otimes N}$ \cite{Fujiwara2003, Kolodynski2013}
yielding the optimal estimation precision $\Delta^2 \eta = \frac{1-\eta^2}{N}$
Let us also note here, that in the  limit of large $N$, simple classes of one- and two-axis spin-squeezed states
reaches the optimal phase estimation precision limit given by $\Delta^2 \varphi = \frac{1-\eta^2}{\eta^2 N} $ \cite{Orgikh2001, Escher2011}.


Due to the high degree of symmetry, it is convenient to shift to angular momentum notation. In general, we write $|j,m\rangle$ to denote a general angular momentum eigenstate where for $N$ qubits $0\leq j\leq \frac{N}{2}$ and $j$ goes between these limits in integer steps  (with a lower bound of $\frac{1}{2}$ for odd $N$) and similarly  $-j \leq m \leq j$, where $m$ also increases in integer steps. We can then write the permutationally symmetric pure input states as $|\psi\rangle=\sum_m \alpha_m |\frac{N}{2},m\rangle$, where $\sum_m|\alpha_m|^2=1$.

After experiencing local dephasing the state will no longer be supported on the fully symmetric subspace $j=N/2$ but will
preserve permutational invariance on the level of the density matrix.
A particularly useful construction for the decomposition of the output state of this evolution is found in \cite{Jarzyna}
\begin{equation}
\label{Blockdiag}
\rho_{\varphi\eta}=\sum_j\sum_{m,m'} h(N,j,m,m',\eta) e^{\mathrm{i}\varphi(m'-m)}|j,m\rangle\langle j,m'|,
\end{equation}
where the actual form of $h(N,j,m,m',\eta)$ coefficients is quite involved and we refer the interested reader to
\cite{Jarzyna} as it has no relevance for further discussion here.
In the above expression it is implicitly assumed that the state has the same form on all multiplicity subspaces corresponding to the same $j$
and we write the state using a simplified notation as if there were no multiplicity of spaces with given $j$.



Let us consider state  $\rho_\eta$, which is an output state \textit{before} implementing the phase evolution---this is permitted because the actions of phase and dephasing commute. We first discuss a further simplification on the structure
of $\rho_\eta$. The parity symmetry implies that within each of the blocks of constant $j$, there exists a further splitting according to the irreducible representations of the parity operator.
The parity operator only has one-dimensional irreducible representations corresponding to the trivial and to the alternating representation. The eigenvectors of $\rho_\eta$ can then be chosen to have either even or odd parity. Given the block diagonal structure, the $i^\t{th}$ even parity vector in the $j$ subspace can be expressed as $|\psi'_{\text{even},i}\rangle=\sum_m e^j_{i,m} (|j,m\rangle + |j,-m\rangle)/\sqrt{2}$, where $\sum_m |e^j_{i,m}|^2 =1$. Similarly, all odd parity eigenvectors have the structure $|\psi'_{\text{odd},i}\rangle=\sum_m o^j_{i,m} (|j,m\rangle - |j,-m\rangle)/\sqrt{2}$, where $\sum_m |o^j_{i,m}|^2 =1$.

Now consider the decomposition of the density matrix in terms of such eigenvectors $\rho_\eta=\sum_i p_i |\psi'_i\rangle\langle\psi'_i|$. The unitary phase only serves to alter the eigenstates. Thus, after the phase unitary the density matrix is $\rho_{\eta\varphi}=\sum_i p_i |\psi_i\rangle\langle\psi_i|$, where $|\psi_i\rangle=U(\varphi)|\psi'_i\rangle$. Due to this $\langle\psi_i|\partial_\eta\psi_k\rangle=\langle \psi_i|U^\dagger(\varphi)\partial_\eta (U(\varphi)|\psi_k\rangle)=\langle \psi'_i| \partial_\eta\psi'_k\rangle$. This simplifies the calculation of $\langle \psi_i| \partial_\eta \rho_{\eta \varphi}|\psi_k\rangle$ terms when $i\neq k$. Most significantly, one can observe that for $\psi'_i$ and $\psi'_k$ from subspaces corresponding to different parities, $\langle \psi_i |\partial_\eta\psi_k \rangle=0$.

This is because the subspaces as a whole do not change with $\eta$. Considering the almost-trivial example of the decomposition of two qubits into triplets and singlets, the singlet space always remains completely separate from the triplet space and it will not overlap with any combination of triplets regardless of $\eta$. The parity subspaces behave similarly.

This eliminates approximately half of the terms of equation \eqref{eq:compatibilityexplicit}. We turn our focus to the remaining terms which include $|\psi_i\rangle$ and $|\psi_j\rangle$ from the \textit{same} parity subspace. We will treat the even parity case, but the proof for odd parity is identical. After the phase unitary, the eigenstates become $|\psi_{\text{even},i}\rangle=\sum_m e_{i,m}^j (e^{- \mathrm{i} \frac{\varphi}{2}m}|j,m\rangle+e^{ \mathrm{i} \frac{\varphi}{2}m}|j,-m\rangle)$. Differentiating this state with respect to $\varphi$ induces a sign difference between the two terms sharing the coefficient $e_{i,m}^j$. Using the orthonormality of $|j,m\rangle$ gives us $\langle\psi_{\text{even},i}|\partial_\phi\psi_{\text{even},k}\rangle=\sum_m e^{j,*}_{i,m}e^j_{k,m}(m-m)=0$. Thus every numerator term, $\langle \psi_i|\partial_\eta \rho|\psi_j\rangle\langle\psi_j|\partial_\phi\rho|\psi_i\rangle$, of $\tr \left(\rho_{\eta,\phi}L_\varphi L_\eta\right)$ is equal to 0 and we can simultaneously estimate the parameters.

\begin{figure}[t]
\includegraphics[width=0.9\columnwidth]{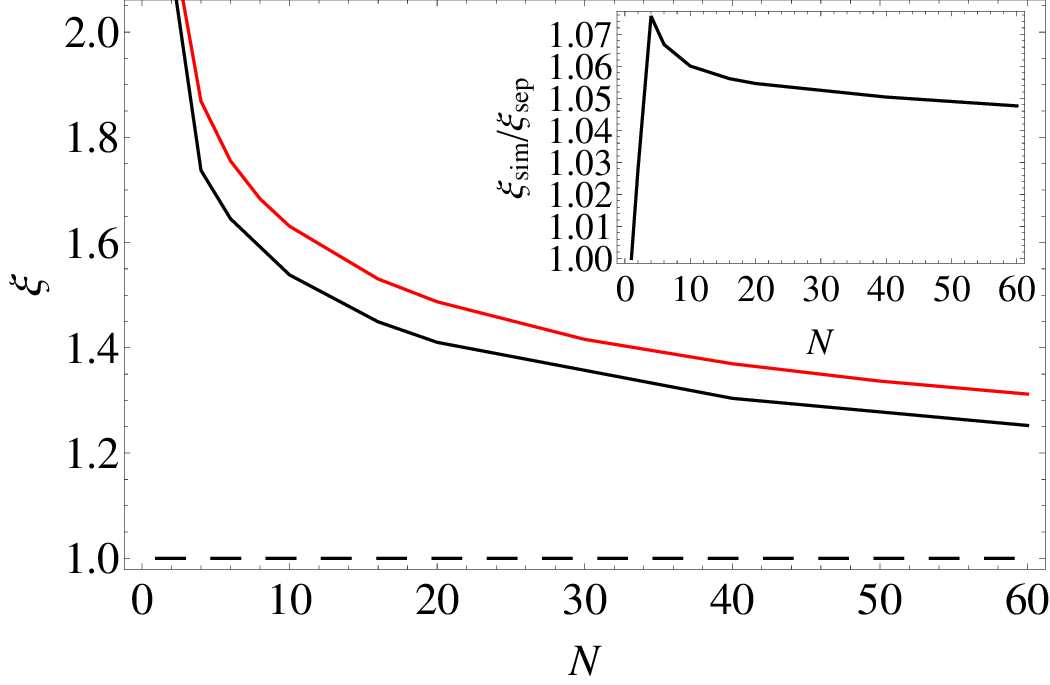}
\caption{Average $\xi$ of normalized uncertainties of estimating the phase and the dephasing parameter in the optimal simultaneous scheme (upper red line) and
the optimal separate schemes (lower black line) as a function of the number of atoms used and the dephasing parameter set to $\eta = 0.9$. For the separate schemes the considered average asymptotically saturates to $1$, which is represented by black dashed line. The inset indicates the ratio of the precision achieved in both schemes indicating that the discrepancy is relatively small and decreases with increasing $N$ which indicates the possibility of satisfying the compatibility requirement in the asymptotic regime of large $N$.}
\label{fig:dephasing}
\end{figure}

There still remains the issue of existence of a common input state optimal  both for $\varphi$ and $\eta$ simultaneously.
This fact is obvious for the single qubit case, $N=1$, as any equatorial qubit state is an optimal probe from the point of view of both parameters. For $N \geq 2$, however, this is no longer true. We have performed a numerical search which showed that
when optimizing probe states from the point of view of estimating two-parameters simultaneously we face a trade-off
 and the optimal state for joint estimation  depends on the weighting of importance between dephasing and phase estimation.
 Still, the observed trade-off is relatively small and shrinks with increasing $N$. We conjecture that for asymptotically large $N$ the discrepancy is vanishing and the simultaneous scheme performs as well as the separate one.
  This is presented in Fig.~\ref{fig:dephasing} where the average of estimation uncertainties of $\varphi$ and $\eta$ achievable when utilizing two-axis spin-squeezed states \cite{Ma2011} normalized according to the asymptotic optimal performance of the separate schemes
 \begin{equation}
  \xi = \frac{1}{2}\left(\frac{\Delta^2\varphi}{(1-\eta^2)/(\eta^2 N)}  +  \frac{\Delta^2\eta}{(1-\eta^2)/N}  \right)
 \end{equation}
is plotted.  When the above quantity is calculated for separate and simultaneous schemes for $\eta=0.9$ the resulting discrepancy is maximal for $N=4$ when it achieves about $7.6 \%$  and decreases with increasing $N$ going below $4.8 \%$ for $N=60$  (see the inset). For smaller $\eta$ the discrepancy is even smaller although its maximum is attained for larger $N$. This numerics strongly suggests that asymptotically simultaneous scheme can perform as well as the separate one.
The two-axis spin squeezed states used as an input probe here are parameterized using the squeezing parameter $\theta$ as
$\ket{\psi_\theta} = e^{-\mathrm{i}\theta\left(J_{+}^2-J_{-}^2\right)}\ket{j,j}$ where $J_{+},\,J_{-}$ are standard angular momentum ladder operators and the  dependence of optimal squeezing parameter as a function of $N$ is approximately $\theta \sim N^{-0.9}$ when $\eta=0.9$. We have also checked the behavior of one-axis spin-squeezed states, recently used in quantum enhanced magnetometry \cite{Muessel2014, Sewell2012}, which are defined as $\ket{\psi_\theta}=U_{\phi}^{SSS}e^{-\mathrm{i}\theta J_x^2}\ket{j,j}$, where $J_x$ is $x$ component of the angular momentum, $U_{\phi}^{SSS}=e^{iH\phi}$ denotes unitary transformation generated by operator $H=e^{\mathrm{i}\theta J_x^2}J_ze^{-\mathrm{i}\theta J_x^2}$ and $\phi=\frac{1}{4}\arctan\left[\frac{4\sin\theta(\cos\theta)^{N-2}}{1-[\cos(2\theta)]^{N-2}}\right]$. Surprisingly, we have found that such states give significantly worse results and do not allow to saturate the performance of the separate schemes.

These conclusions are therefore similar to the ones obtained in \cite{Durkin} where a different model assuming collective instead of uncorrelated dephasing was analyzed and again asymptotic possibility of performing optimal simultanous estimation of phase and the dephasing parameter has been demonstrated.

\section{Conclusions}
\label{sec:conclusions}
We have presented a complete analysis of the compatibility problem in multiparameter quantum metrology, pointing out
three main obstacles to estimating parameters simultaneously with the same accuracy as in the separate scenario. We have provided several examples which illustrate how these obstructions come into force, as well as being interesting in their own right.

We would like to stress, however, that multiparameter metrology is not all about trying to avoid an overwhelming array of pitfalls.
In this paper we have taken the specific approach in which we were asking for a multiparameter protocol to meet the performance
of the separate schemes where each of the parameters is estimated independently with the highest possible precision possible.
Clearly, even if a multiparameter scheme cannot meet this condition, it does not mean that there is no advantage in estimating multiple parameters simultaneously.  In general, there will be an advantage coming from simultaneous estimation
 even if the compatibility conditions are not satisfied. This has indeed been the line of research of many other papers dealing with multiparameter metrology.
 From this point of view, one can view this paper as providing a systematic view on the situation when multiparameter estimation manifests its maximal advantage over separate schemes by  meeting their performance while consuming a factor of $p$ fewer resources.

It is also interesting to comment on the issue of sequential vs. parallel schemes in quantum metrology in
the multiparameter case. It is known that in decoherence-free single unitary parameter estimation
a scheme where unitaries act sequentially on a single probe provides the same maximal QFI as the parallel scheme where
one allows arbitrary input entangled state of $N$ particles to be sent through $N$ parallel unitaries \cite{GioMac} and only the presence of decoherence makes the schemes inequivalent \cite{Demkowicz2014}.
We have shown that using two-qubit entangled input states allows one to optimally estimate two rotation angles
around perpendicular axes with precision equal to that which could be obtained in the separate scheme.
Clearly, this could not be achieved by acting sequentially with two unitaries on a single qubit as in this case we have proven
that the compatibility condition cannot be satisfied when two parameters are to be estimated. This breaks the equivalence between entangled and sequential unitary parameter estimation in the multiparameter case.



\begin{acknowledgements}
We thank Madalin  Gu{\c{t}}{\u{a}} for pointing out to us the importance of QLAN in the context of the problem considered in this paper. We thank S. Zhou and L. Jiang for notifying us of an error in Eq.~\eqref{eq:boundholevo} of an earlier version of the paper. SR is supported by EPSRC's Quantum Communications Hub.
\end{acknowledgements}
\bibliography{CompatibleMetrology}
\end{document}